\documentclass{PoS}
\usepackage{bm}
\newcommand{\Jpsi}{J\!/\!\psi}

\title{Production of \bm{$J\!/\!\psi$} quarkonia in color evaporation model based on \bm{$k_{T}$}-factorization}

\ShortTitle{$J\!/\!\psi$ in color evaporation model}

\author{\speaker{Rafa{\l} Maciu{\l}a}\thanks{This study was partially supported by the Polish National Science Center
grant DEC-2014/15/B/ST2/02528 and by the Center for Innovation and
Transfer of Natural Sciences and Engineering Knowledge in Rzesz{\'o}w.}\\
        Institute of Nuclear
Physics, Polish Academy of Sciences, Radzikowskiego 152, PL-31-342 Krak{\'o}w, Poland\\
        E-mail: \email{rafal.maciula@ifj.edu.pl}}

\author{Antoni Szczurek\\
        Institute of Nuclear
Physics, Polish Academy of Sciences, Radzikowskiego 152, PL-31-342 Krak{\'o}w, Poland\\
        E-mail: \email{antoni.szczurek@ifj.edu.pl}}

\abstract{We use a new approach to color evaporation model (CEM)
for quarkonium production.
The production of $c\bar c$ pairs is performed within 
$k_T$-factorization approach using different unintegrated gluon
distribution functions (UGDF) from the literature.  
We include all recent improvements to color evaporation model.
We get poor description of the large transverse momentum distributions
of $J/\psi$ with the JH-2013 CCFM-based UGDF. Here explicit
inclusion of $2 \to 3$ processes considerably improves the situation.
Similar effects are discussed in the context of the KMR UGDF.}

\FullConference{XXVII International Workshop on Deep-Inelastic Scattering and Related Subjects - DIS2019\\
		8-12 April, 2019\\
		Torino, Italy}

\begin{document}

\section{Introduction}

Inclusive production of quarkonia is one of the most actively studied
topics at the LHC. The $\Jpsi$, $\Psi'$, $\Upsilon$, $\Upsilon'$
and $\Upsilon''$ are the usually measured quarkonia.
The production of $\Jpsi$ is a model case. There was (still is) a
disagreement related to the underlying production mechanism.
There are essentially two approaches. The first one is the so-called
nonrelativistic QCD (NRQCD) approach (see \textit{e.g.} Ref.~\cite{Chang:1979nn}). 
There are two versions of such an approach based on collinear 
or $k_T$-factorization approaches. 
It was shown recently that the LHC data can be explained
within the NRQCD $k_T$-factorization approach with a reasonable set of
parameters \cite{CS2018}. 

Another popular approach is color evaporation model (CEM) \cite{Fritsch,Halzen}.
In this approach one is using perturbative calculation of $c \bar c$.
The $c \bar c$-pair by emitting a soft radiation goes to color singlet
state of a given spin and parity. The emission is not explicit in this
approach and everything is contained in a suitable renormalization
of the $c \bar c$ cross section when integrating over certain limits
in the $c \bar c$ invariant mass. It was proposed recently how to extend
the original CEM and improved color evaporation model (ICEM) was developed (see \textit{e.g.} Ref.~\cite{MV2016}).
Usually the computations of the transverse momentum dependence of $\Jpsi$ meson production within the color evaporation model are based on collinear approach up to the next-to-leading
order (NLO).
Within the collinear-factorization approach in the leading-order (LO) approximation transverse momentum of the $c\bar c$-pair is equal to zero.
In fact, the NLO diagrams for inclusive $c$-quark, such as 
$gg \to g c \bar c$ or $qg \to q c \bar c$, constitute the leading-order contributions for the $c\bar c$-pair transverse momentum. Similarly, the next-to-next-to-leading-order (NNLO) topologies for this quantity are effectively NLO.
The situation is different in the $k_{T}$-factorization approach where non-zero $c\bar c$-pair transverse momentum
can be obtained already at leading-order within the $g^*g^* \to c \bar c$ or $q^*\bar{q}^* \to c \bar c$ mechanisms.

It was shown many times, that the $k_{T}$-factorization with the KMR unintegrated
distributions turned out to be successful in the description of
inclusive production of $D$ mesons \cite{Maciula:2013wg,Maciula:2018iuh}
as well as for some correlation observables 
\cite{Maciula:2013wg,Karpishkov:2016hnx} at the LHC. 
It seems therefore interesting, and valueable,
to apply the $k_{T}$-factorization approach for $c \bar c$ production 
in the context of the color evaporation model for $\Jpsi$ meson 
production (see also Ref.~\cite{Cheung:2018tvq}).
In the present paper we wish to study whether such a combination
of elements can allow to describe the world data for prompt
$\Jpsi$ production.

%--------------------------------------------------------------
\section{Theoretical framework}
%--------------------------------------------------------------

\begin{figure}[!h]
\centering
\begin{minipage}{0.25\textwidth}
  \centerline{\includegraphics[width=1.0\textwidth]{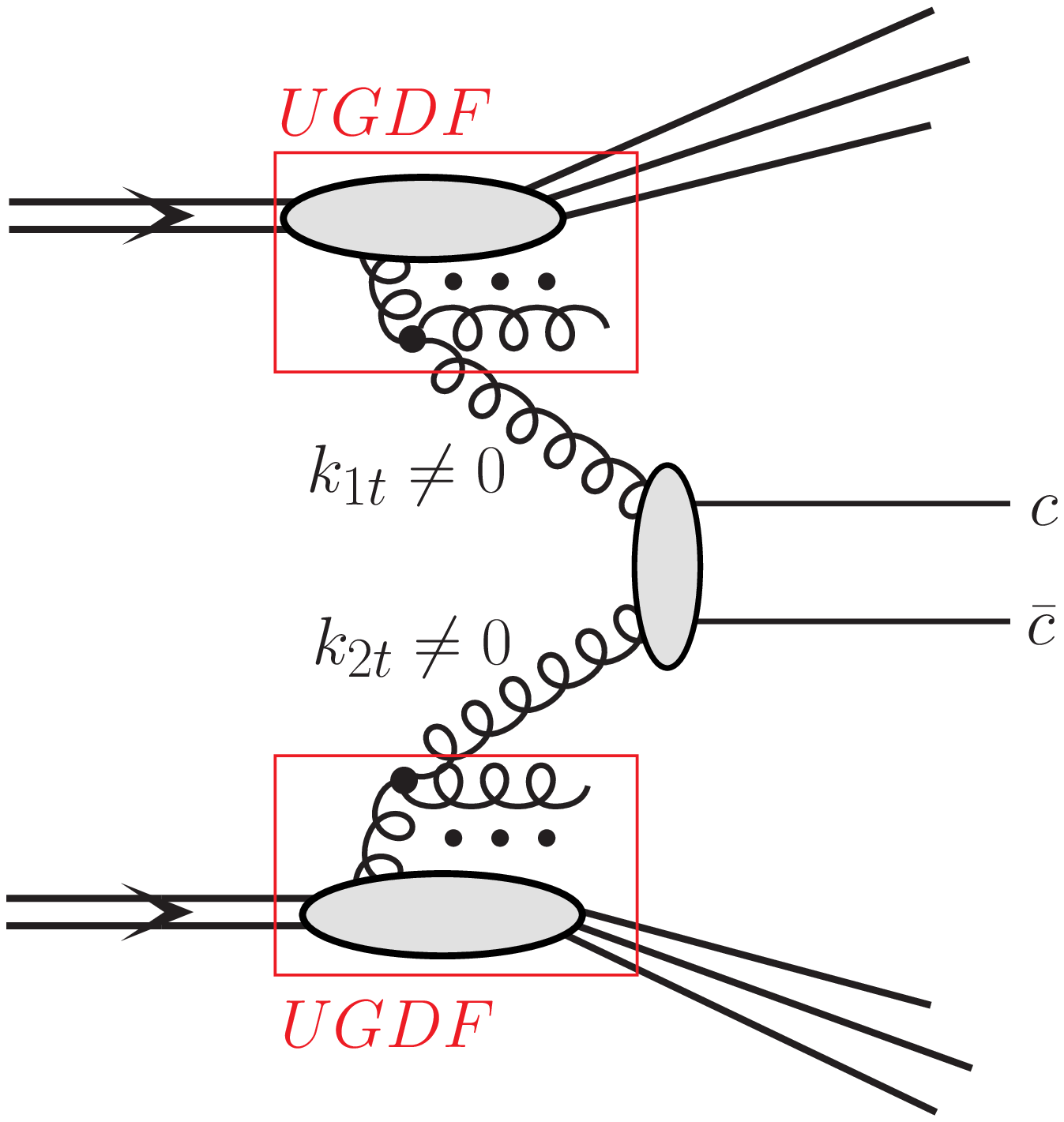}}
\end{minipage}
\begin{minipage}{0.25\textwidth}
  \centerline{\includegraphics[width=1.0\textwidth]{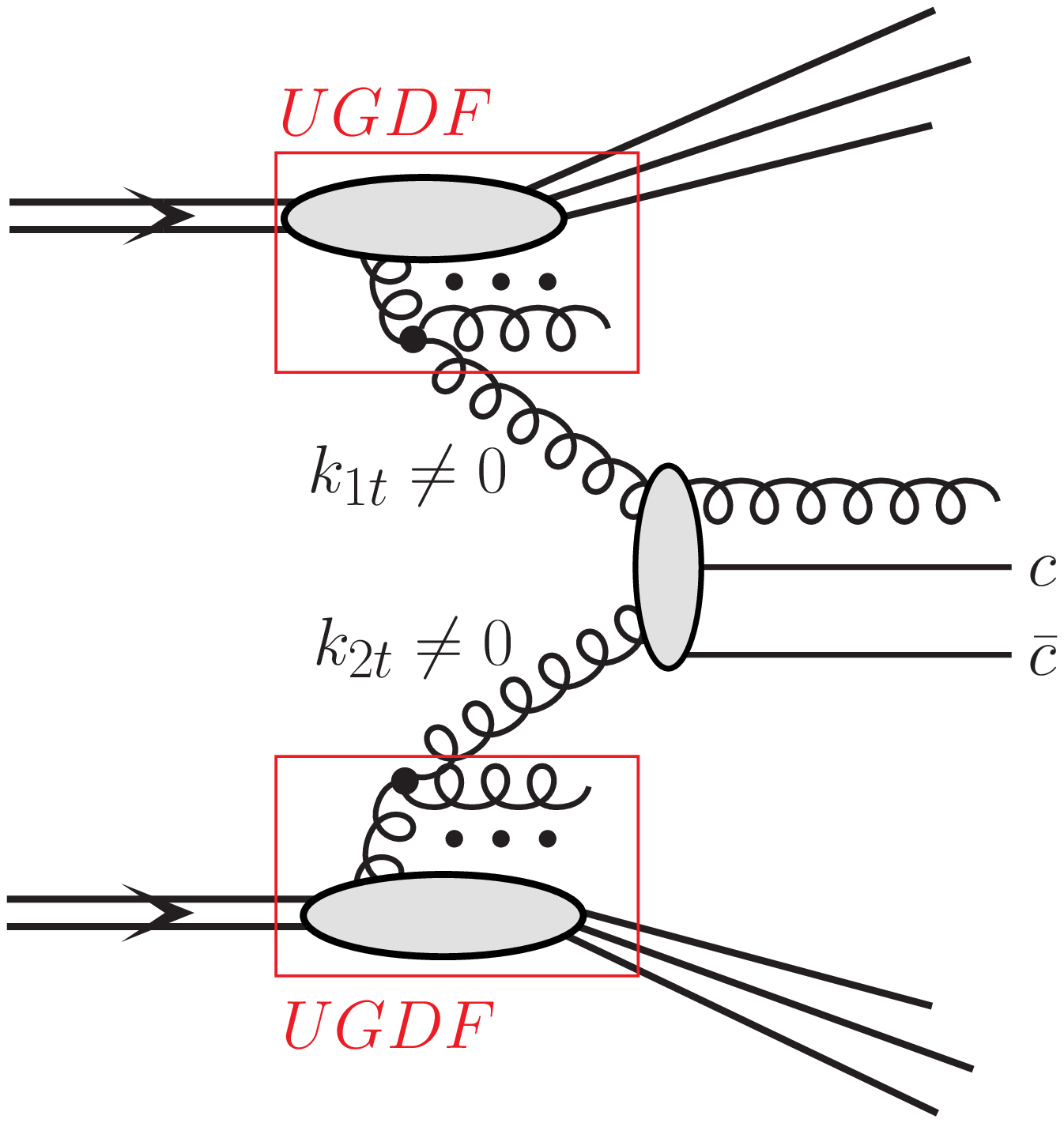}}
\end{minipage}
  \caption{
\small A diagrammatic representation of the $g^*g^* \rightarrow c\bar c$ (left) and $g^*g^* \rightarrow g c\bar c$ (right) mechanisms under consideration.
}
\label{fig:diagrams}
\end{figure}

In the basic step of our approach, i.e. calculation of the cross section for $c\bar c$-pair production, we follow the $k_{T}$-factorization approach. This framework was shown many times by different authors to provide very good description of heavy quark production in proton-proton collisions at different energies. In principle, the $k_{T}$-factorization approach is known to be a very efficient framework for studies of correlation observables, such as $c\bar c$ invariant mass or $c\bar c$-pair transverse momentum distributions \cite{Maciula:2013wg,Karpishkov:2016hnx}. Within the CEM the transverse momentum distribution of $\Jpsi$ meson is strictly connected to the $c\bar c$-pair transverse momentum. In the collinear approach, the sole graphs contributing to the production of the $c\bar c$-pair with a finite pair transverse momentum
are those from $2 \to 3$ processes.
In the $k_{T}$-factorization approach also the contributions from $2 \to 2$ mechanisms become available.
In the present paper, we take into consideration both dominant components: $g^{*}g^{*} \to c \bar c$ and $g^{*}g^{*} \to g c \bar c$ , shown schematically in the left and right panels of Fig.~\ref{fig:diagrams}, respectively.

%----------------------------------------------------------------------------
%----------------------------------------------------------------------------

According to this approach, the transverse momenta (virtualities) of both partons entering the hard process are taken into account and the sum of transverse momenta of the final $c$ and $\bar c$ no longer cancels. Then the differential cross section at the tree-level for the $c \bar c$-pair production reads:
\begin{eqnarray}\label{LO_kt-factorization} 
\frac{d \sigma(p p \to c \bar c \, X)}{d y_1 d y_2 d^2p_{1,t} d^2p_{2,t}} &=&
\int \frac{d^2 k_{1,t}}{\pi} \frac{d^2 k_{2,t}}{\pi}
\frac{1}{16 \pi^2 (x_1 x_2 s)^2} \; \overline{ | {\cal M}^{\mathrm{off-shell}}_{g^* g^* \to c \bar c} |^2}
 \\  
&& \times  \; \delta^{2} \left( \vec{k}_{1,t} + \vec{k}_{2,t} 
                 - \vec{p}_{1,t} - \vec{p}_{2,t} \right) \;
{\cal F}_g(x_1,k_{1,t}^2) \; {\cal F}_g(x_2,k_{2,t}^2) \; \nonumber ,   
\end{eqnarray}
where ${\cal F}_g(x_1,k_{1,t}^2)$ and ${\cal F}_g(x_2,k_{2,t}^2)$
are the unintegrated gluon distribution functions (UGDFs) for both colliding hadrons and ${\cal M}^{\mathrm{off-shell}}_{g^* g^* \to c \bar c}$ is the off-shell matrix element for the hard subprocess. The extra integration is over transverse momenta of the initial
partons. The matrix element squared for off-shell gluons is taken here in the
analytic form proposed by Catani, Ciafaloni and Hautmann (CCH)
\cite{Catani:1990eg}.

The framework of the $k_{T}$-factorization is also used in the calculation of the $g^*g^* \to g c\bar c$ hard subprocess.
The off-shell matrix elements for higher final state parton multiplicities at the tree level can be calculated e.g.
numerically with the help of methods of numerical BCFW recursion. The calculations are performed with the help of \textsc{KaTie} package \cite{vanHameren:2016kkz}, which is a complete Monte Carlo parton-level event generator for
hadron scattering processes. At tree-level the relevant calculations of the $g^*g^* \to g c\bar c$
contribution can be performed with a special treatment of minijets at
low transverse momenta by multiplying standard cross section by a
somewhat arbitrary suppression factor $F_{sup}(p_T) =
\frac{p_T^4}{((p_{T}^{0})^{2} + p_T^2)^2}$. This method for the
regularization of the cross section is also applied in the \textsc{Pythia} Monte
Carlo generator for light quark and gluon $2\to
2$ partonic processes. There, the free parameter $p_{T}^{0}$ of the
suppression factor is fitted to the experimental data on total cross
section. In our case, the default value of the free parameter is
$p_{T}^{0} = 1.5$ GeV which is adjusted to the exact NLO calculations of
the charm production cross section at the LHC.

In the numerical calculation below we apply the
Kimber-Martin-Ryskin (KMR) unintegrated gluon distributions \cite{Watt:2003mx} that was found recently to work very well in the case of charm production at the LHC \cite{Maciula:2013wg}.
For completeness of the present studies, we also use the CCFM-based
JH-2013 UGDFs \cite{JH2013} that were applied in the same
context in Ref.~\cite{Cheung:2018tvq}.

The KMR prescription for UGDF allows for contributions from the region of $k_{t} > \mu_{F}$. In other words, contains additional hard emissions from the gluon ladder. As a consequence, calculating the $g^*g^* \rightarrow c\bar c$ mechanism with the KMR UGDFs one effectively includes part of higher-order diagrams with one and even two associated partons. Other models of UGDFs in the literature usually contain only soft emissions in the ladder and omit the $k_{t} > \mu_{F}$ region, which is the case of \textit{e.g.} the JH-2013 UGDFs. Within the latter case, the contributions with associated minijets or jets should be included rather in an explicit way. This difference in the construction of the UGDFs may be crucial especially for correlation observables, such as the pair transverse momentum.

Having calculated differential cross section for $c\bar c$-pair
production one can obtain the cross section for $J\!/\!\psi$-meson
within the framework of ICEM \cite{MV2016}. The $c\bar c \to J\!/\!\psi$ transition can be
formally written as follows:
\begin{eqnarray}
\frac{d\sigma_{J\!/\!\psi}(P_{J\!/\!\psi})}{d^3P_{J\!/\!\psi}} &= F_{J\!/\!\psi} \int_{M_{J\!/\!\psi}}^{2M_D} d^3 P_{c\bar c} \; d M_{c\bar c} \frac{d\sigma_{c\bar c}(M_{c\bar c},P_{c\bar c})}{ d M_{c\bar c} d^3 P_{c\bar c}} \delta^3(\vec{P}_{J\!/\!\psi}-\frac{M_{J\!/\!\psi}}{M_{c\bar c}} \vec{P}_{c \bar c}),
\end{eqnarray}
where  $F_{J\!/\!\psi}$ is the probability of the $c\bar c \to J\!/\!\psi$ transition which is fitted to the experimental data, $M_{J\!/\!\psi}$ (or $M_{D}$) is the mass of $J\!/\!\psi$ (or $D$) meson and $M_{c\bar c}$ is the invariant mass of the $c\bar c$-system.
Using the momentum relation
\begin{equation}
\vec{P}_{J\!/\!\psi}=\frac{M_{J\!/\!\psi}}{M_{c\bar c}} \vec{P}_{c \bar c}, \;\; \mathrm{where} \;\; \vec{P}_{c \bar c} = \vec{p}_{c} + \vec{p}_{\bar c}, 
\end{equation}
one can easily calculate also rapidity of $J\!/\!\psi$-meson. In the last step, in order to compare the theoretical predictions with the prompt $\Jpsi$ experimental data we correct the numerical results by the direct-to-prompt ratio equal to $0.62$ \cite{Digal:2001ue}.  

\section{Numerical results}

%----------------------------------------------------------------------------
\begin{figure}[!h]
\centering
\begin{minipage}{0.37\textwidth}
  \centerline{\includegraphics[width=1.0\textwidth]{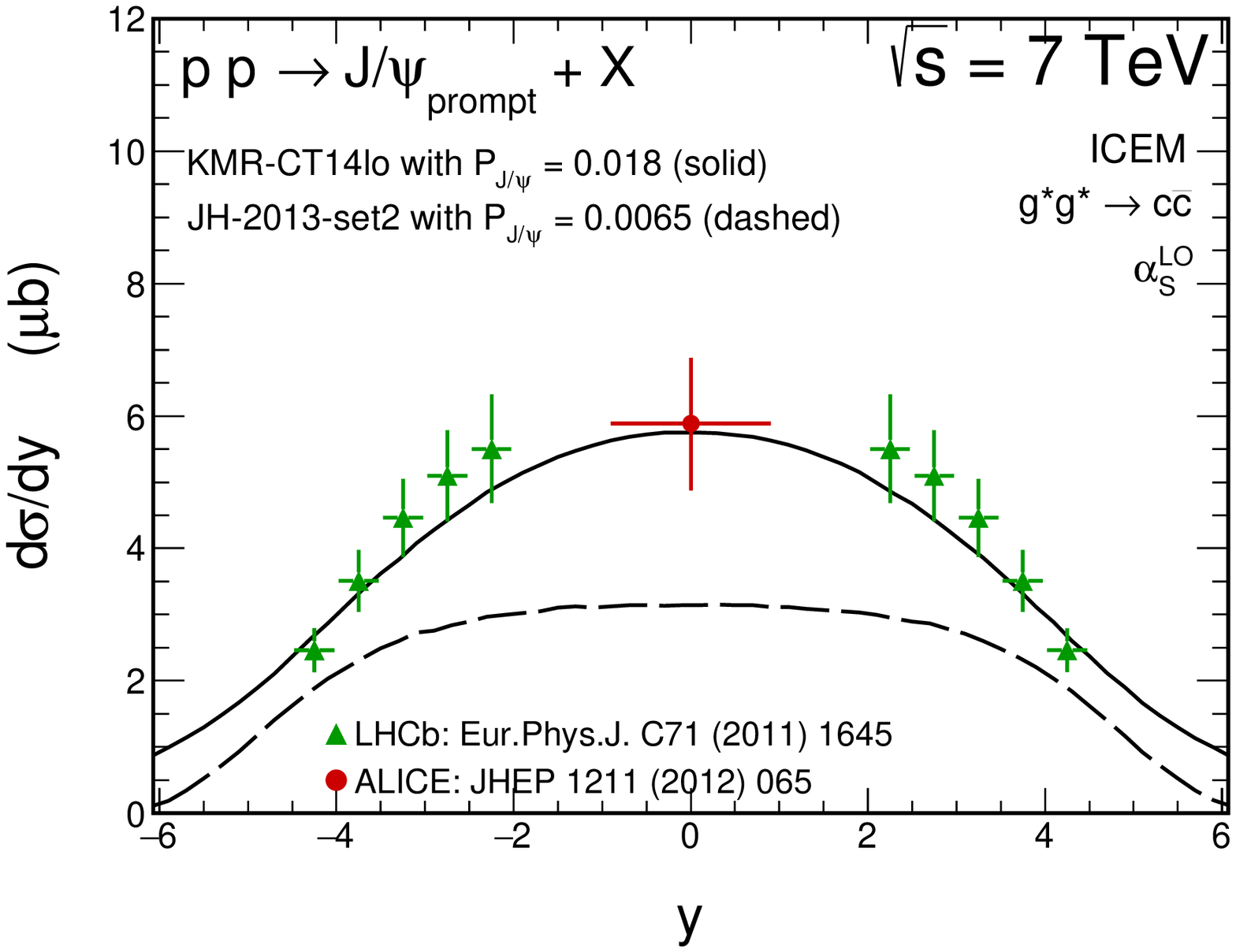}}
\end{minipage}
\begin{minipage}{0.37\textwidth}
  \centerline{\includegraphics[width=1.0\textwidth]{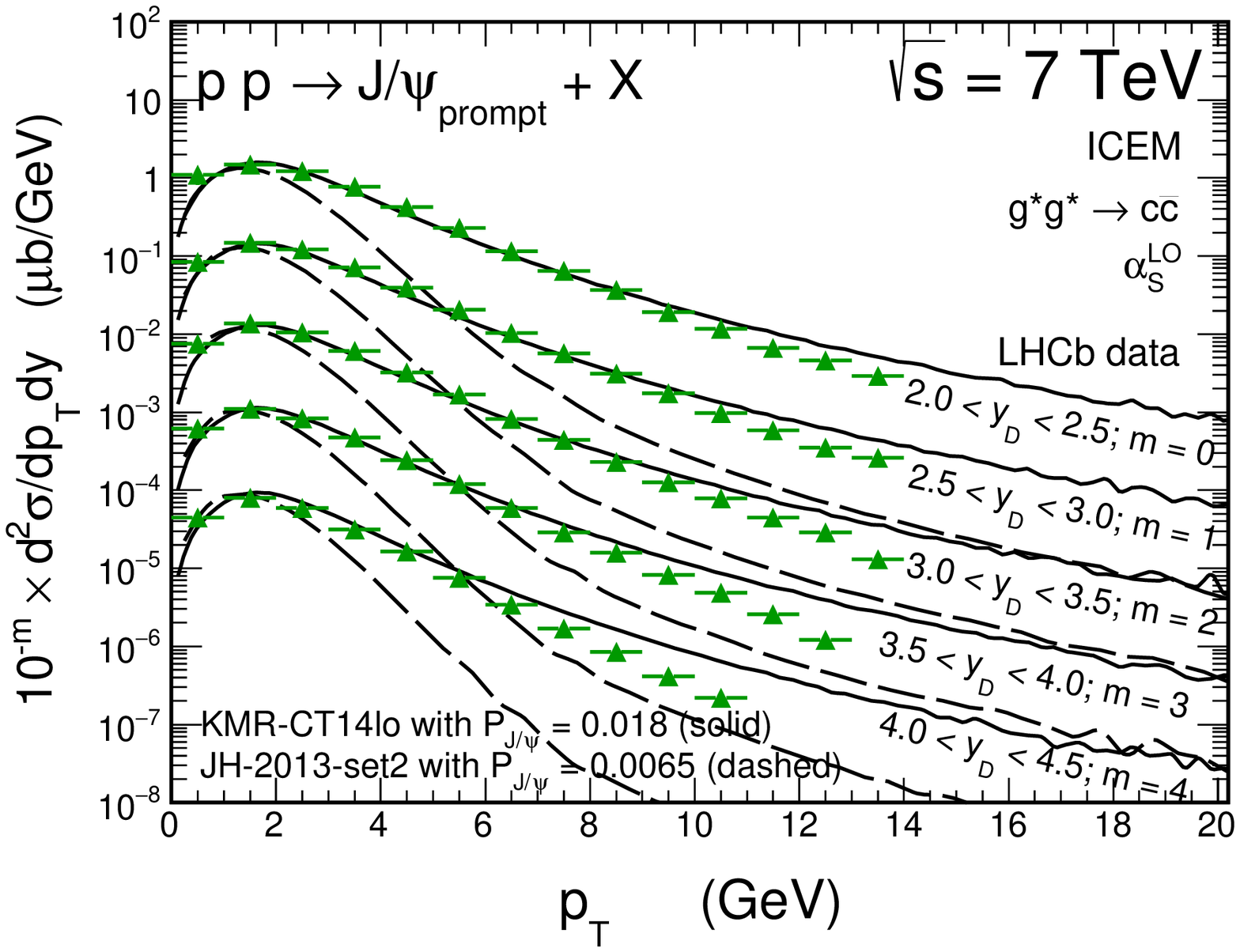}}
\end{minipage}
  \caption{
\small Distributions in rapidity and transverse momentum of prompt $\Jpsi$ 
for $\sqrt{s}$ = 7 TeV obtained within the $k_{T}$-factorization realization
of the ICEM for different UGDFs. The LHCb data were measured only on one
side of $y = 0$. We have added them symmetrically on the other side
as often done in the literature.
}
\label{fig:JPsi_meson_1}
\end{figure}
%----------------------------------------------------------------------------

In the left and right panel of Fig.~\ref{fig:JPsi_meson_1} we show the $\Jpsi$-meson rapidity and transverse momentum distributions, respectively, together with the
ALICE \cite{Abelev:2012gx} and the LHCb data \cite{Aaij:2011jh}. Here we present results for the default KMR-CT14lo (solid lines) and for the JH-2013-set2 (dashed lines) UGDFs. In this calculation the model parameter $P_{\Jpsi}$ was fixed to 0.018, and to 0.0065, respectively, in order to describe the LHCb data at small transverse momenta. Within these values, we get a very good description of the experimental rapidity distribution of the prompt $\Jpsi$-meson in the case of the KMR-CT14lo UGDF. 
In principle, one could fit both rapidity distributions with
the same quality taking different values of $P_{\Jpsi}$. However, the
calculated transverse momentum distributions differ strongly and
the LHCb data prefers the result with the KMR-CT14lo UGDF.
For the JH-2013-set2 UGDF, the larger values of the $P_{\Jpsi}$ suggested by the rapidity spectrum would lead to a significant overestimation of the small transverse momentum data.
The distributions obtained with this UGDF have completely different $p_{T}$-slope than the experimental one and falls down much faster. This observation is consistent with the results presented in Ref.~\cite{Cheung:2018tvq}. There, this behavior of the quarkonium $p_{T}$-distributions was, in our opinion, correctly recognized as a consequence of omitting of the $k_{t} > \mu$ region in the UGDF. As mentioned in the previous section only the KMR model includes this contribution explicitly. In the case of the JH-2013 UGDF additional emissions of an extra hard gluon can be taken into account by 
exact calculations of the $g^*g^* \to g c \bar c$ mechanism at the level of matrix elements. 

%----------------------------------------------------------------------------
\begin{figure}[!h]
\centering
\begin{minipage}{0.37\textwidth}
  \centerline{\includegraphics[width=1.0\textwidth]{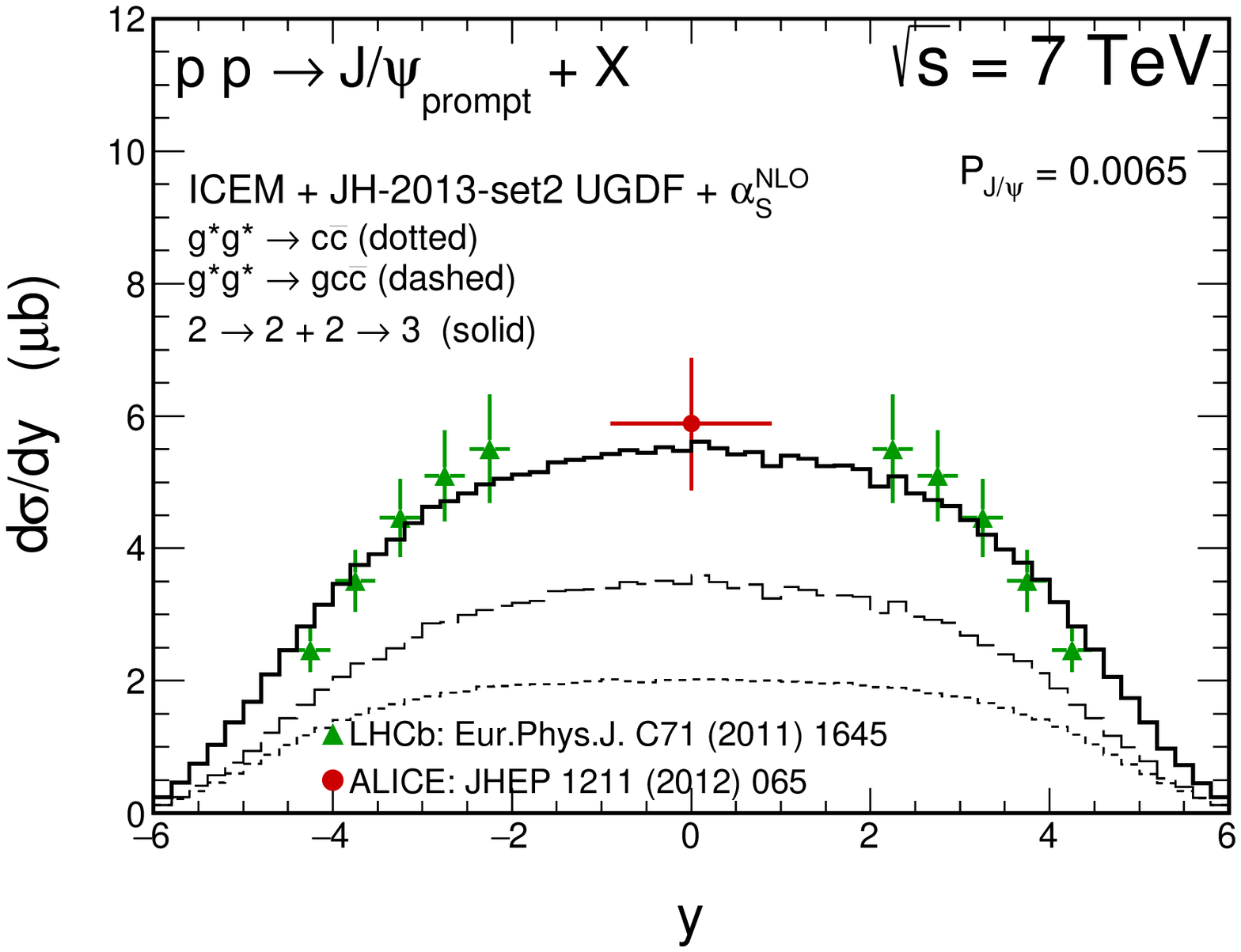}}
\end{minipage}
\begin{minipage}{0.37\textwidth}
  \centerline{\includegraphics[width=1.0\textwidth]{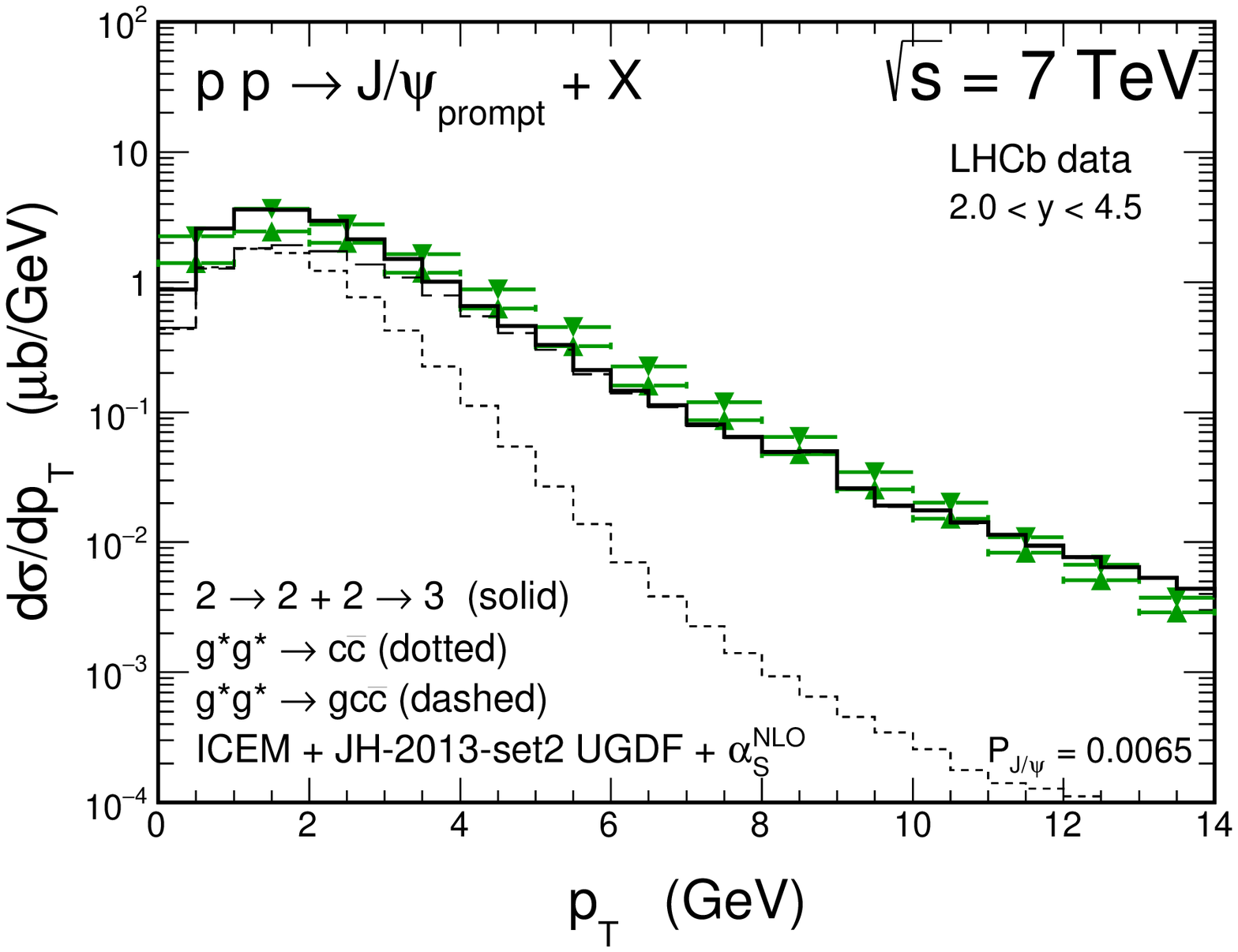}}
\end{minipage}
\begin{minipage}{0.37\textwidth}
  \centerline{\includegraphics[width=1.0\textwidth]{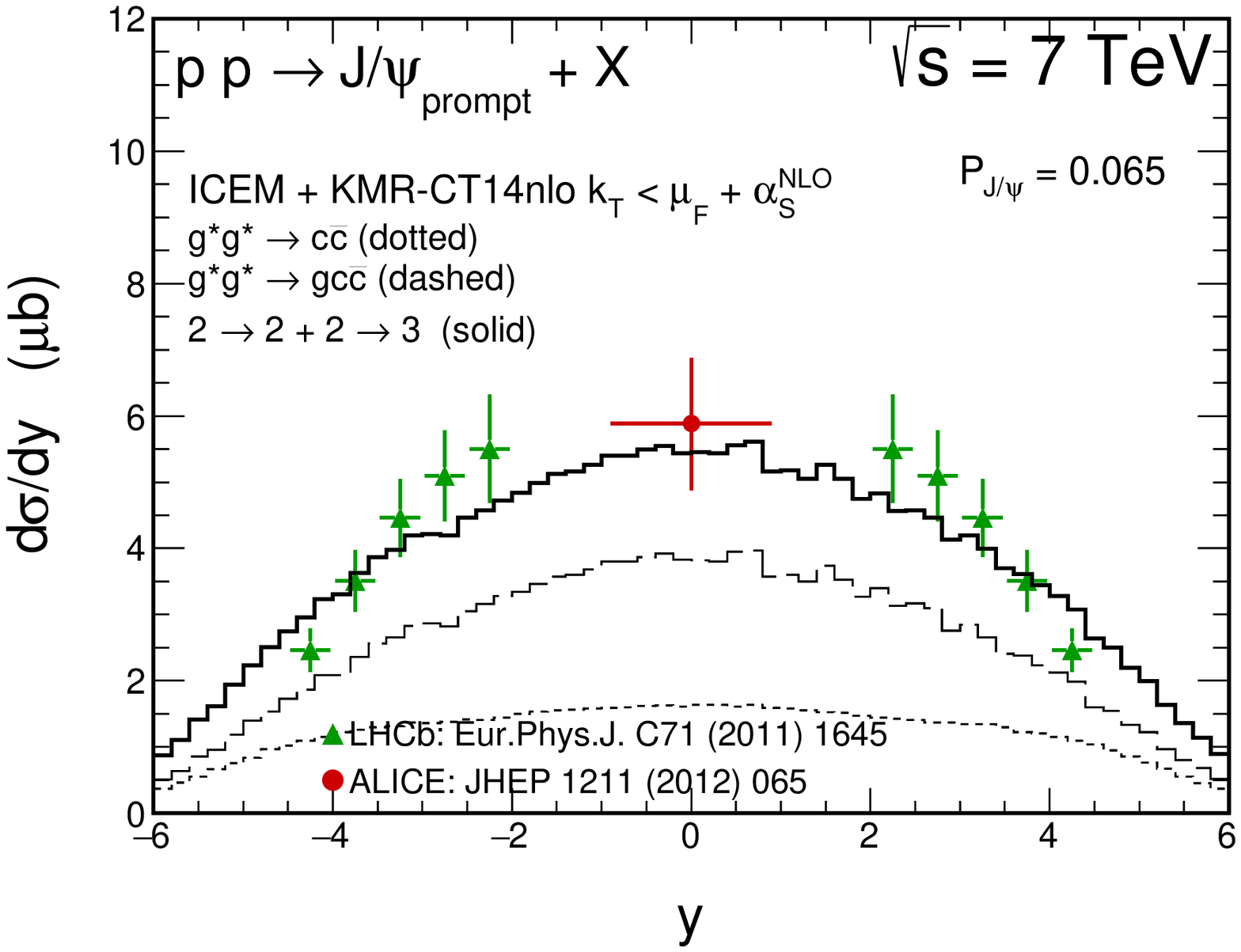}}
\end{minipage}
\begin{minipage}{0.37\textwidth}
  \centerline{\includegraphics[width=1.0\textwidth]{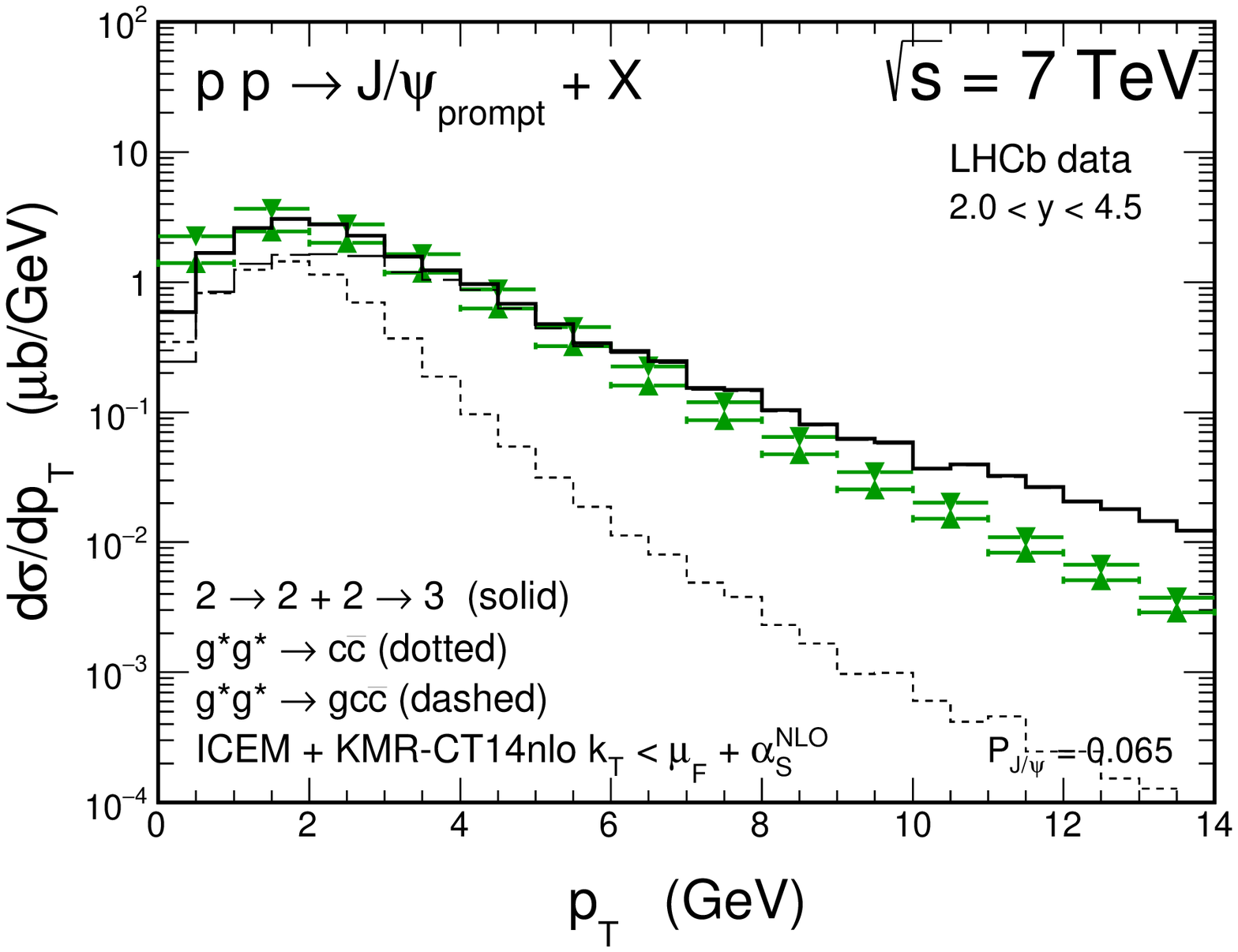}}
\end{minipage}
  \caption{
\small Distributions in rapidity and transverse momentum of prompt $\Jpsi$
for $\sqrt{s}$ = 7 TeV obtained within the $k_{T}$-factorization realization
of the ICEM for different UGDFs. Here, both the $g^*g^* \to c\bar c$ and the $g^*g^* \to g c\bar c$ mechanisms
are taken into account.
}
\label{fig:JPsi_meson_2to3}
\end{figure}
%----------------------------------------------------------------------------

It was shown in Ref.~\cite{Maciula:2013wg} that the KMR and CCFM-based UGDFs lead to significant differences in correlation observables for $c\bar c$-pair, \textit{e.g.} in $D\bar{D}$ invariant mass and/or azimuthal angle distributions. In the case of the CCFM-based unintegrated gluon distributions, an improved description of correlation observables
can be obtained once the higher-order process of gluon-splitting is taken into account in explicit way \cite{Jung:2010ey}. 
The JH-2013 model for UGDFs by its construction does not include the
contributions from the $k_{t} > \mu_{F}$ region and resume
only soft extra emissions. Therefore, working within this model we find
necessary to include contributions with additional hard (mini)jets at
the level of hard matrix elements. Having regard to the lack of the NLO
framework for the $k_{T}$-factorization, this can be done by adding
together the $g^*g^* \to c\bar c$ and $g^*g^* \to g c\bar c$
mechanisms. At high energies, the $2 \to 3$ channel driven by the
gluon-gluon fusion is the dominant one. In
Fig.~\ref{fig:JPsi_meson_2to3} we present numerical results obtained
within this procedure. Here, 
for consistency, the running coupling constant $\alpha_{S}$ is taken at next-to-leading order, in contrast to the previous calculations. The top and bottom panels correspond to the JH-2013-set2 and the KMR UGDFs, respecitvely. The dotted histograms correspond to the
$g^*g^* \to c\bar c$ mechanism and the dashed histograms are for the $g^*g^* \to g c\bar c$.  
In the case of the KMR predictions, here we impose a special cut $k_{t} < \mu_{F}$ to avoid double counting.  
We see that the presence of the $g^*g^* \to g c\bar c$ mechanism completely changes the picture for the JH-2013-set2 UGDF
and allows for a very good description of the experimental data. The proposed procedure does not influence the predictions
for the KMR UGDF. Both considered prescriptions ($2\to 2$
with the standard KMR approach and $2\to 2 + 2\to 3$ with the modified
KMR approach) provide a description of the experimental data of a similar quality.   

%--------------------------
\section{Conclusions}
%--------------------------

In the present paper we have discussed how to extend the improved 
color evaporation model for production of $\Jpsi$ meson to be used in the framework of 
$k_{T}$-factorization approach for production of $c$ and $\bar c$ pairs.
The same was done independently very recently in Ref.~\cite{Cheung:2018tvq}.
We have included recent developments proposed recently in the literature.
In our calculations we have used the KMR unintegrated gluon 
distributions which allows to relatively well describe 
the single $D$-meson distributions as well as meson correlation observables.

The CCFM-based JH-2013 UGDF leads to a rather poor agreement with 
the LHCb transverse momentum distributions at large $p_T$.
Much better agreement is achieved when including explicitly extra
emissions of (mini)jets. In the case of the KMR approach it is sufficient to use the standard
$k_T$-factorization approach and allow for initial gluon $k_T$ larger
than factorization scales. A corresponding discussion has been presented.

\end{document}